\documentclass[a4paper,10pt]{paper}

\usepackage{latexsym}
\usepackage{amsmath}
\usepackage{amssymb,amsfonts}
\usepackage{graphicx}
\usepackage{cite}
\usepackage{bm,wrapfig}   
\usepackage{upgreek}

\numberwithin{equation}{section}
\def\theequation{\arabic{section}.\arabic{equation}}

\def\beq{\begin{eqnarray}}
\def\eeq{\end{eqnarray}}
\def\T{\tau}

\def\R{{\hbox{{\rm I}\kern-.2em\hbox{\rm R}}}} 
\def\H{{\hbox{{\rm I}\kern-.2em\hbox{\rm H}}}} 
\def\N{{\hbox{{\rm I}\kern-.2em\hbox{\rm N}}}} 
\def\C{{\ \hbox{{\rm I}\kern-.6em\hbox{\bf C}}}} 
\def\Z{{\hbox{{\rm Z}\kern-.4em\hbox{\rm Z}}}} 
 
\newcommand{\ve}{\mathbf}

\newcommand{\ack}[1]{\par\section*{Acknowledgement} #1}

\catcode`\@=11
\@addtoreset{equation}{section}

\begin{document}

\title{Unruh--DeWitt detectors in spherically symmetric dynamical space-times} 

\author{G. Acquaviva\thanks{acquaviva@science.unitn.it} \and
R.~Di~Criscienzo\thanks{rdicris@science.unitn.it}
\and M. Tolotti
\and L.~Vanzo\thanks{vanzo@science.unitn.it}
\and S.~Zerbini\thanks{zerbini@science.unitn.it}
\\
\\
Dipartimento di Fisica - Universit\`a di Trento 
\\
and Istituto Nazionale di Fisica Nucleare,
\\
Via Sommarive 14, 38123 Povo, Italia}


\maketitle

\begin{abstract}
In the present paper, Unruh--DeWitt detectors are used in order to investigate the issue of temperature associated with a spherically symmetric dynamical space-times. Firstly, we review  the semi-classical tunneling method, then we introduce the Unruh--DeWitt detector approach. We show that for the generic static black hole case and the FRW de Sitter case, making use of peculiar Kodama trajectories, semiclassical and quantum field theoretic techniques give the same standard and well known thermal interpretation, with an associated temperature, corrected by appropriate Tolman factors. For a FRW space-time interpolating de Sitter space with the Einstein--de Sitter universe (that is a more realistic situation in the frame of $\Lambda$CDM cosmologies),  we show that the detector response splits into a de Sitter contribution plus a fluctuating term containing no trace of Boltzmann-like factors, but rather describing the way thermal equilibrium is reached in the late time limit. As a consequence, and unlike the case of black holes, the identification of the dynamical surface gravity of a cosmological trapping horizon as an effective temperature parameter seems lost, at least for our co-moving simplified detectors. The possibility remains that a detector performing a proper motion along a Kodama trajectory may register something more, in which case the horizon surface gravity would be associated more likely to vacuum correlations than to particle creation.
\end{abstract}


\section{Introduction}

It is well known that Hawking radiation \cite{Haw} is considered one of the most important predictions of quantum field theory in 
curved space-time. Several derivations of this effect have been proposed \cite{dewitt,BD,wald,fulling,igor} and recently the search for ``experimental'' verification making use of analogue models has been pursued by many investigators (see for example \cite{Unruh:2008zz,Barcelo:2005fc}). One of the most beautiful achievements is represented by the interpretation of the surface gravity $\kappa$ associated to a (black hole or cosmological) horizon as the (Hawking) temperature associated to the radiation emitted from that horizon: a quantum effect as explicitly shown by the famous formula $T_H = \hbar c^3 \kappa/2\pi k_B$. As long as we limit to consider stationary Killing horizons all the derivations of this formula are basically equivalent; but, going from stationary to dynamical geometries, things make tough mostly due to the lack of a time translation symmetry generator. Limiting to spherical symmetric geometries, it is possible to show \cite{kodama} the existence of a vector field which, defined through the condition $(\mathcal K^\alpha G_{\alpha \beta})^{; \beta} =0$, resembles most of the amazing Killing vector properties. We shall term this vector the \textit{Kodama vector field} and refer the interested reader to \cite{bob07,sean09,bob09,av} for more accurate discussion. By choosing Kodama observers as privileged and using semi-classical techniques, we can still establish a correspondence between surface gravity and temperature in dynamical, spherical symmetric, black hole or cosmological space-times. Semi-classical methods such as the tunneling method or the Hamilton--Jacobi method have proven so far to be reliable in all the testable conditions (for a recent review on the subject \cite{tr} and references therein). However, a comparison between such methods and standard quantum field theoretic calculations in dynamical spaces is still something deserving to be done. With the purpose of partially filling this gap, we try to understand how far we can push the correspondence between surface gravity and temperature in dynamical spherically symmetric spaces. We shall do this by searching whether a point-like detector will register a quasi-thermal excitation rate of the form
$$
\dot{F}\sim E\exp(-2\pi E/\kappa_{H}(t))
$$
where $\kappa_{H}(t)$ is the dynamical horizon surface gravity to be introduced below, and $E$ is the energy of the detector's quantum jump. We shall take this feature as a hint that the detector feels the vacuum as a mixed quasi-thermal state with an effective temperature parameter $T_{H}(t)=\kappa_{H}(t)/2\pi$. The analysis will be done for conformally coupled scalar fields, the best approximation we know to massless radiation, but we think the conclusions to be drawn will have general validity. 
Important examples will cover black holes, de Sitter space and generic FRW space-times.

The paper is organized as follows. In  Section 2, we briefly review basic predictions of the tunneling method. Section 3 contains a discussion on the Unruh--DeWitt detector which is introduced in order to confirm through quantum field theoretic calculations the results of previous tunneling papers. In Section 4 we show how the formalism of the previous section applies to a generic static black hole, and finite time effects are investigated. Section 5 is devoted to FRW space-times, and to a discussion of the realistic model of a universe filled with matter and cosmological constant $(\Omega_{m},\Omega_{\Lambda})$, where the main conclusions are presented as consequence of analytic results supported by  numerical computations.  Concluding remarks follow at the end of the paper.

We use the metric signature $(-,+,+,+)$; Greek indices run over $0$ to $3$ while mid-Latin as $i,j$ only over $0$ and $1$. We use Planck units in which $c=\hbar=G=k_B=1$. 

\section{The Hamilton--Jacobi method}

In 2000 Parikh and Wilczek \cite{PW} (see also \cite{visser}) introduced the so-called tunneling approach for investigating Hawking radiation. Here, we shall focus on a variant of their method, called Hamilton--Jacobi tunneling method \cite{Angh,tanu,mann,obr}. This method is covariant and can be extended to the dynamical case 
\cite{bob07,bob08,sean09,bob09}, and to the study of decay of massive particles and particle creation by naked singularities \cite{bob10}. 
In their approach, Parikh and Wilczek made a clever use of the Painlev\'e--Gullstrand stationary gauge for four-dimensional Schwarzschild black hole 
\beq
ds^2 &= & -\left(1-\frac{2 M}{r}\right)d t_p^2 + 2\sqrt{\frac{2 M}{r}}dr dt_p+dr^2+ \nonumber \\
& & \hspace{1cm} + \;r^2 (d\theta^2 + \sin^2\theta d\phi^2)
\eeq
which is regular on the trapping horizon $r_H=2M\;$ \footnote{The global event horizon is $r_0<r_H$ for an evaporating black hole and $r_{0}>r_{H}$ for an accreting black hole, as can be seen from the equation of radial null rays, $\dot{r}_{0}=1-\sqrt{r_{H}/r_{0}}$.}. This is one of the key points since the use of singular gauges, as the Schwarzschild gauge, leads, in general, to ambiguities and it is useless in the dynamical case which we are interested in.
The second merit we can address to Parikh and Wilczek work is the treatment of back-reaction on the metric, based on energy conservation. In the following, we shall limit to leading term results and neglect the issue of back-reaction. However, it may be worth to recall that in the limit where the number of emitted quanta is reasonably large back-reaction effects can be accounted for by assuming that the mass parameter $M$ is a continuous function of time $t_p$. Penrose's diagrams for this more general case have been determined too, {\it e.g.} in \cite{Lindesay:2010uv,Brown:2009by}.

The Hamilton--Jacobi method is reasonably simple, even though subtleties are present (see, for example \cite{seanrep}). It is based on the computation of the classical action $I$ along a trajectory starting slightly behind the trapping horizon but ending in the bulk, and the associated WKB approximation (keeping $\hbar$ explicit)
\beq
\mbox{Amplitude} \propto e^{i \frac{I}{\hbar}}\, .
\eeq 
For an evaporating black hole such a trajectory would be classically forbidden since at the trapping horizon photons are only momentarily at rest, whence $dr/dt_p<0$ inside the horizon\footnote{It means the photon must go back in time $t_p$ to escape the horizon.}. The related semiclassical emission rate reads 
\beq
\Gamma \propto |\mbox{Amplitude}|^2 \propto 
e^{-2\frac{\Im I}{\hbar}} \,. 
\eeq
with $\Im$ standing for the imaginary part. In the tunneling across the horizon, the 
imaginary part of the classical action $I$ stems from the interpretation of a formal horizon divergence, and in evaluating it, one has to make use of Feynman's prescription related to a simple pole in integration.
This corresponds to the choice of suitable boundary conditions in quantum field theory approach.

It should be mentioned also that in the static/stationary case, there exist  different interpretations to the Hamilton--Jacobi method (see, for example \cite{D}), none of them, however, can be easily extended to the dynamical case, the one we are mainly interested in.

We may anticipate that in the WKB approximation of the tunneling probability, one asymptotically gets a Boltzmann factor,
\beq
\Gamma \propto e^{- \frac{\beta}{\hbar} \omega}\,,
\eeq
where $\omega$ represents the Kodama energy of the tunneling particle, $\omega= - \mathcal K \cdot d I$. 
It is crucial in our approach that the argument of the exponent be a coordinate scalar (invariant quantity), since otherwise no physical meaning can be addressed to $\Gamma$. In particular, in the Boltzmann factor, $\beta$ and the energy $\omega$ have to be separately coordinate scalars, otherwise again no invariant meaning could be given to the quantity $\beta/\hbar$. In the static case, we interpret $T=\hbar/\beta$ as the horizon temperature. 
The  tunneling method for spherically symmetric dynamical case, has been considered in \cite{sean09,bob09,kodama,sean,peng}, where the Kodama--Hayward invariant formalism has been used.

In the cosmological case on the other hand, one can still have a trapping horizon despite the absence of collapsed matter simply as a result of the expansion of the universe. As for an excreting black hole, this too is represented by a time-like hypersurface. Similarly, an approximate notion of temperature can be associated to such horizons based on the existence of a surface gravity and again the tunneling method gives a non vanishing amplitude having the Boltzmann form. However, we will see that a co-moving monopole detector seems to react to the expansion in a different, non ``Boltzmannian'' way, while reaching thermal equilibrium (or better, detailed balance conditions) only asymptotically for large times. 

As showed in \cite{bob09}, in a generic FRW space-time, one has 
\beq
\Gamma \sim \exp\left(-2\Im \,I\right) \sim e^{-\frac{2 \pi}{(-\kappa_H)}\, \omega_H}\;,
\eeq
where, since $\kappa_H <0$ and $\omega_H >0$ for physical particles, the imaginary part of the action is positive definite. This result is invariant since the quantities appearing in the imaginary part are manifestly invariant. Furthermore $T=-\kappa_H/2 \pi$ satisfies a First Law.
As a consequence, at least in some asymptotic regime and for slowly changes in the geometry, one could
interpret $T=-\kappa_H/2 \pi$ as the dynamical temperature associated with FRW space-times. 
In particular, this gives naturally a positive temperature for de Sitter space-time, a long debated question years ago, usually resolved by changing the sign of the horizon's energy.  It should be noted that in literature, the dynamical temperature is usually given in the form $T=H/2\pi$ (exceptions are the papers \cite{Wu:2008ir,chen}) with $H^{2}=\Lambda/3\equiv H_{0}^{2}$. Of course this is the expected result for de Sitter space in inflationary coordinates, but it ceases to be correct in any other coordinate system. 
This fact seems not so widely known so, for sake of completeness, we shall try to show it in detail. de Sitter space in the global patch is described by the metric
\beq
ds^2=-dt^2+ a^2(t) d\Omega_{(3)}^2 \;
\label{dsfrwnf} 
\eeq
with $a(t)=\cosh(H_0t)$, $d\Omega_{(3)}^2$ the unit three-sphere and $\hat k=H_0^2$ is the only relevant scale. The Hubble parameter is time dependent, $H(t)=H_0\tanh (H_0t)$, and satisfies the identity $\dot H(t)=H_0^2-H^2(t)=\hat k/a^2(t)$.
The horizon radius turns out to be
\beq 
R_H:= a(t) r_H = \frac{1}{\sqrt{ H^2(t)+\frac{\hat k}{a^2(t)}}}=\frac{1}{H_0}
\label{h0}
\eeq 
and Hayward's surface gravity (minus sign due to conventions) is
\beq
 \kappa_H = - \left( H^2(t) +\frac{1}{2} \dot H(t) + \frac{\hat k}{2a^2(t)}\right) R_H(t) =H_0
\label{hfrw0} 
\eeq
an invariant quantity, indeed. Hence, we see that the $\dot H$ and $\hat k$ terms have to be present in a generic
FRW space-time. The important spatially flat case straightforwardly follows, $\kappa_H= - [ H(t) +\dot H(t)/2 H(t)]$.
Note that this is independent on position, suggesting that $\kappa_{H}$ really is an intrinsic property of FRW space linked to the bulk. The horizon's temperature and the ensuing heating of matter was foreseen several years ago in the interesting paper 
\cite{Brout:1987tq}.

\section{The Unruh--DeWitt detector}

We recall that for the decay rate of a massive particle in de Sitter space, the exact quantum field theory calculation of Moschella \textit{et al.}\ \cite{ugo} supports the WKB semiclassical tunneling result of \cite{Volo,bob10}. 

What about the energy scale associated with the horizon tunneling? We have shown that 
the semiclassical WKB method leads to an asymptotic particle production rate, involving the Boltzmann factor and related ``temperature''
\beq
\Gamma \simeq e^{-\frac{2 \pi}{|\kappa_H|} \omega_H}\quad \rightarrow T=\frac{|\kappa_H|}{2\pi}
\eeq
where $\kappa_H$ is the Hayward invariant surface gravity. For a generic spherically symmetric space-time, this result obtained by the Hamilton--Jacobi tunneling method seems a very clear prediction, namely an answer to the question: how hot is our expanding universe? 
A possible way to understand this issue using quantum field theory in curved space-time is to make use of a ``quantum thermometer'' (basically, a Unruh--DeWitt detector) and evaluate its response function, that is, loosely speaking, the number of clicks per unit proper time it detects as it is carried around the universe. For a recent review, see \cite{crispino}. 

In our approach, since we would like to obtain an invariant result, we will consider detectors which follow Kodama trajectories in a generic spherically symmetric space-time. The problem of back-reaction on these Kodama trajectories has been investigated in \cite{casadio}.

As we will see, the Unruh--DeWitt thermometer gives a clean answer only in the stationary case, while for FRW space-time the situation is not so simple, since horizon effects are entangled with highly non trivial kinematic effects. For general trajectories in flat space-time see the recent paper \cite{Kothawala:2009aj}. An interesting analysis has been also put forward by Obadia \cite{obadia08}. 
In a recent paper \cite{nico}, local scaling limit techniques have been used in investigating the Hawking radiation.

In the following, we review the well know Unruh--DeWitt detector formalism, adapted to a spherically symmetric 
conformally flat space-time, namely, introducing the conformal time $\eta$ by means of $d\eta=\frac{d t}{a}$, 
the flat FRW space-time we are going to deal with reads  
\begin{equation}
ds^2=a^2(\eta)(-d\eta^2+d \ve x^2)\,,\quad \qquad x=(\eta, \ve x)\;.
\label{cf} 
\end{equation}
For the purpose at hand, it is very convenient to consider  a free massless scalar field which is conformally coupled to gravity, since, as is well known,  the related Wightman function $W(x,x')$ can be computed in an exact way. In fact, one has 
\begin{equation}
W(x,x')=\sum_{\ve k} f_{\ve k}(x)f^*_{\ve k}(x')\,,
\label{w} 
\end{equation}
where the mode functions $f_{\ve k}(x)$ satisfy the conformally invariant equation ($\mathcal R$ being the curvature scalar)
\begin{equation}
\left(\Box -\frac{\mathcal R}{6}\right) f_{\ve k}(x)=0\;.
\label{f} 
\end{equation}
Making the ansatz 
\begin{equation}
f_{\ve k}(x)=\frac{g(\eta)}{a(\eta)}e^{-i\ve k \cdot \ve x}\;,
\label{g} 
\end{equation}
one has, for the unknown quantity $g(k)$,
\begin{equation}
g''(\eta) + k^2 g(\eta) =0\;.
\label{g1} 
\end{equation}
As a consequence, making the choice of the vacuum given by
\beq
g(\eta)=\frac{e^{-i\eta \vert k\vert }}{2\sqrt{\vert k\vert}}\,,
\eeq
one  has  \cite{BD} 
\begin{equation}
W(x,x')= \frac{1}{a(\eta)a(\eta')}\,\frac{1/4\pi^2}{|\ve x- \ve x'|^2-|\eta-\eta'-i\epsilon|^2 }\,.
\label{wold} 
\end{equation}
As is usual in distribution theory we shall leave understood the limit as $\epsilon\to 0^{+}$.
However, it has been shown by Takagi \cite{T} and Schlicht \cite{S} that this prescription is manifestly non-covariant. 
Since one is dealing with distributions, the limit $\epsilon \rightarrow 0^+$ has to be taken in the weak sense, and
it may lead to unphysical results with regard to instantaneous proper-time rate in Minkowski space-time. We adapt Schlicht's proposal to our conformally flat case, namely
\begin{equation}
W(x,x')=\frac{1}{a(\eta)a(\eta')}\,\frac{1/4\pi^2}{[( x- x')-i\epsilon (\dot{x}+\dot{x}')]^2 }\,.
\label{w3} 
\end{equation}
where an over dot stands for derivative with respect to proper time 
In the flat case, this result has been generalized by Milgrom and Obadia, who made use of an analytical proper-time regularization \cite{langlois,OM}. It should be noted the appearance of Minkowski contribution, as a function of the conformal time $\eta$. 

The transition probability per unit proper time of the detector depends on the response function per unit proper time which, for radial trajectories, at finite time $\tau$ may be written as  \cite{louko} 
\beq 
& &\frac{d F}{ d\tau}=\frac{1}{2\pi^2}\mbox{Re}\int_{0}^{\tau -\tau_0} ds
\frac{e^{-i E s}}{a(\tau)a(\tau-s)} \times\nonumber \\
& &  \frac{1}{[x(\tau)-x(\tau-s)- i\epsilon(\dot{x}(\tau)+\dot{x}(\tau-s))]^2}
\eeq
where $\tau_0$ is the detector's proper time at which we turn on the detector, and $E$ is the energy associated with the excited detector state (we are considering $E>0$).  
Although the covariant $i\epsilon$-prescription is necessary in order to deal with the second order pole at $s=0$, one may try to avoid the awkward limit $\epsilon \rightarrow 0^+$ by omitting the $\epsilon$-terms but subtracting the leading pole at $s=0$ (see \cite{louko} for details). In fact, the normalization condition
\beq
g_{\mu\nu}\dot{x}^{\mu}\dot{x}^{\nu}\equiv\left[a(\tau) \dot x(\tau)\right]^2= -1\, ,
\label{normaliz}
\eeq
characteristic of time-like four-velocities, has to be imposed. Thus, introducing the notation
\beq
\sigma^2(\tau,s) \equiv a(\tau)a(\tau-s)[x(\tau)-x(\tau-s)]^2\,, 
\eeq
due to (\ref{normaliz}), for small $s$, one has
\beq
\sigma^2(\tau,s)=-s^2[1+s^2d(\tau,s)]\,.
\eeq
As a consequence, for $\Delta \tau = \tau-\tau_0 >0$, one can present the detector transition probability per unit time in the form
\beq 
\frac{d F}{ d\tau}=\frac{1}{2 \pi^2}\int_{0}^{\infty}ds\, \cos( E s)
\left(\frac{1}{\sigma^2(\tau,s)} + \frac{1}{s^2}\right) + J_\tau
\label{L6}
\eeq 
where the ``tail'' or finite time fluctuating term is given by
\beq
J_\tau := -\frac{1}{2 \pi^2}\int_{\Delta \tau}^{\infty}ds\,\frac{\cos(E\,s)}{\sigma^2(\tau,s)}\,.
\label{L00}
\eeq
The convergence at infinity is assumed, but in all physically interesting cases it is ensured. This is not quite the original expression found in \cite{louko} but can be obtained from it by simple manipulations.
This is the main formula which we will use in the following.  Equation \eqref{L6} may be  much more convenient to deal with than the original expression containing the $\epsilon$-terms, since in the latter the limit in distributional sense must also be taken at the end of any computation.\\
In the important stationary case in which $\sigma(\tau,s)^2=\sigma^2(s)=\sigma^2(-s)$, Eq.~\eqref{L6} simply becomes
\beq 
\frac{d F}{ d\tau}=\frac{1}{4\pi^2}\,\int_{-\infty}^{\infty}ds\, e^{-i E s}\left( \frac{1}{\sigma^2(s)}+\frac{1}{s^2}\right) + J_\tau\,. \label{Lr}
\eeq 
In this case (examples are the static black hole and the FRW de Sitter space) all the finite time dependence is contained in the fluctuating tail.\\

As a result we have the  manageable expression \eqref{L6}, in which the last fluctuating tail term incorporates part of the finite-time effects and, as we will see, in the case of asymptotically stationary situations controls how fast the thermal equilibrium is reached.

\section{Quantum thermometers in static and stationary spaces}

As an application of the formalism previously developed, first we are going to revisit the case of the generic static black hole, then we shall consider the perturbation of thermal equilibrium due to finite-time effects.  

\subsection{The generic static black hole}

The general metric for a static black hole reads 
\beq
ds^2=-V(r)dt^2+\frac{dr^2}{W(r)}+r^2 d \Omega^2 \;,
\label{s} 
\eeq
where, for sake of simplicity, we shall assume $W(r)=V(r)$, with $V(r)$ having just simple poles in order to describe what we might call a {\it nice black hole} (suggested by Hayward). Let $r_H$ be the (greatest) solution of $V(r)=0$, the general formalism tells us that the horizon is located at $r=r_H$; the Kodama vector coincides with the usual Killing vector $(1,\ve 0)$; and the Hayward surface gravity is the Killing surface gravity, namely $\kappa_H=\kappa=V'_H/2$. 
We now introduce the Kruskal-like gauge associated with this static black hole solution. The first step consists in introducing the tortoise coordinate
\beq
r^*(r)=\int^r\frac{d\tilde r}{V(\tilde r)}\,.
\eeq
One has $-\infty < r^*< \infty$ and
\beq
ds^2
&=&V(r^*)[-dt^2+(dr^{*})^2]+r^2(r^*) d \Omega^2_{(2)} \;.
\label{s*} 
\eeq
The Kruskal-like coordinates are
\beq
R=\frac{1}{\kappa}e^{\kappa r^*}\cosh(\kappa t)\,, \quad T=\frac{1}{\kappa}e^{\kappa r^*}\sinh(\kappa t)\,,
\eeq
so that
\beq
-T^2+R^2=\frac{1}{\kappa^2}e^{2 \kappa r^*}\,,
\label{ku}
\eeq
and the line element becomes
\beq
ds^2 &=& e^{-2\kappa r^*}\,V(r^*)[-dT^2+dR^2]+r^2(T,R) d \Omega^2 \nonumber \\
&\equiv& e^{\Psi(r^*)}(-dT^2+dR^2)+r^2(T,R) d \Omega^2 
\label{ks*} 
\eeq
where now the coordinates are $T$ and $R$, $r^*=r^*(T,R)$, $e^{\Psi(r*)}:=V(r^*)e^{-2\kappa r^*}$.

The key point to recall here is that in the Kruskal gauge (\ref{ks*}) the normal metric -- the important one for radial trajectories -- is conformally related to two dimensional Minkoswki space-time. The second observation is that Kodama observers are defined by the integral curves associated with the Kodama vector, thus the areal radius $r(T,R)$ and $r^*$ are {\it constant}. As a consequence, the proper time along Kodama trajectories reads
\beq
d\tau^2=V(r^*)dt^2 &=&e^{\Psi(r^*)}(dT^2-dR^2)\nonumber \\  &=& a^2(r^*)(dT^2-dR^2)\
\eeq
so that $t=\tau/\sqrt{V(r^*)}$ and 
\beq
R(\tau)&=&\frac{1}{\kappa}e^{\kappa r^*}\cosh\left(\kappa\frac{\tau}{\sqrt{V(r^*)}}\right) \nonumber\\T(\tau)&=&\frac{1}{\kappa}e^{\kappa r^*}\sinh \left(\kappa \frac{\tau}{\sqrt{V(r^*)}} \right)\,.\label{cazzo}
\eeq 
The geodesic distance reads
\beq
\sigma^2(\tau,s)&=&e^{\Psi(r^*)}\left[ -\left(T(\tau)-T(\tau-s)\right)^2+\right. \nonumber \\
& & \left. +\hspace{10pt} \left(R(\tau)-R(\tau-s)\right)^2 \right]\,,
\eeq
and one gets, using (\ref{cazzo}),
\beq\label{figa}
\sigma^2(\tau,s) &=& -\frac{4 V(r^*)}{\kappa^2} \sinh^2\left(\frac{\kappa\,s}{2\sqrt{V(r^*)}}\right)\,. \label{sinh2}
\eeq
Since $\sigma^2(\tau,s)= \sigma^2(s) = \sigma^2(-s)$, we can use equation (\ref{Lr}) in the limit when $\Delta\tau $ goes to infinity:
\begin{equation} 
\frac{d F}{ d\tau}= \frac{\kappa}{8\pi^2 \sqrt{V^*}}\int_{-\infty}^{\infty}dx e^{- \frac{2 i\sqrt{V*} E x}{\kappa}}\left[ -\frac{1}{\sinh^2 x}+\frac{1}{x^2}\right]\;.
\end{equation}
The integral can be evaluated by the theorem of residues and the final result is
\beq 
\frac{d F}{ d\tau}= \frac{1}{2\pi}\; \frac{E}{\exp\left(\frac{2\pi \sqrt{V^*} E}{\kappa}\right)-1}\,.
\eeq
Since the transition rate exhibits the characteristic Planck distribution, it means that the Unruh--DeWitt thermometer 
in the generic spherically symmetric black hole space-time 
detects a quantum system in thermal equilibrium at the local temperature 
\beq
T=\frac{\kappa}{2\pi \sqrt{V^*}}\,.
\label{z}
\eeq
With regard to the factor $\sqrt{V^*}=\sqrt{-g_{00}}$, recall Tolman's theorem which states that, for a gravitational system at thermal equilibrium, {$T\sqrt{-g_{00}}=\mbox{constant} $. For asymptotically flat black hole space-times, one obtains the ``intrinsic'' constant temperature of the Hawking effect, i.e.
\beq
T_H=\frac{\kappa}{2\pi}= \frac{V'_H}{4\pi}\,.
\eeq

We would like to point out that this is a quite general result, valid for a large class of nice black holes, as for example
Reissner-Nordstr\"{o}m and Schwarzschild-AdS black holes. On the other hand, the Schwarzschild-dS black hole cannot be included, due to the presence of two horizons. However, as an important particular case, we may consider the static patch of de Sitter space, with a metric defined by
\beq
V(r)=1-H_0^2r^2\,, \qquad H_0^2=\frac{\Lambda}{3}\,.
\label{j}
\eeq
The unique horizon is located at $r_H=H_0^{-1}$ and the Gibbons--Hawking temperature is \cite{GH} $T_H=H_0/2\pi$. In the next Section, we will present a derivation of this well known result in another gauge.\\

We conclude this subsection making some remarks on de Sitter and anti-de Sitter black holes. First, we observe that in a static space-time, namely the one corresponding to a nice black hole, the Killing--Kodama observers with $r=K$ constant have an invariant acceleration 
\beq
A^2=g_{\mu\nu}A^\mu A^\nu=\frac{V'^2(K)}{4 V(K)}\,,
\label{a}
\eeq
where $A^\mu = u^\nu \nabla_{\nu} u^\mu$, $u^\mu$ being the observer's four-velocity, that is the (normalized) tangent vector to the integral curves of the Kodama vector field.  In the case of de Sitter black hole, one has
\beq
A^2=\frac{H_0^4 K^2}{1-H_0^2 K^2}\,.
\eeq
As a result, 
\beq
A^2+H_0^2=\frac{H_0^2}{1-H_0^2 K^2}\,,
\eeq
and the de Sitter local temperature felt by the Unruh detector, 
\beq
T_{dS}=\frac{H_0}{2\pi}\frac{1}{\sqrt{1-H_0^2 K^2}}
\eeq
can be re-written as \cite{thirr,deser98}
\beq
T_{dS}=\frac{1}{2\pi}\sqrt{A^2+H_0^2}=\sqrt{T^2_U+T^2_{GH}}\,.
\eeq
A similar result was also obtained for AdS in \cite{deser98}, and it reads
\beq
T_{AdS}=\frac{1}{2\pi}\sqrt{A^2-H_0^2}\,.
\label{d}
\eeq
We would like to show that it is a particular case of our general formula (\ref{z}). In fact, it is sufficient to 
apply it to the four-dimensional topological black hole with hyperbolic horizon manifold found in 
\cite{Beng,Mann,Brill,Vanzo}, which is a nice black hole with
\beq
V(r)=-1-\frac{C}{r}+H_0^2r^2\,,
\eeq
where $C$ is a constant of integration related to the black hole mass. The space-time is a solution of Einstein equation with negative cosmological constant $\Lambda=-3H^2_0$, which is asymptotically Anti-de Sitter. When the constant of 
integration goes to zero, one has still a black hole solution, and calculation similar to the one valid for de Sitter space-time gives
\beq
T_{AdS}=\frac{H_0}{2\pi}\frac{1}{\sqrt{-1+H_0^2 K^2}}=\frac{1}{2\pi}\sqrt{A^2-H_0^2}\,,
\eeq 
which is Deser \textit{et al.} result \cite{deser98}. Thus, for spherically symmetric space-times with constant curvature one has that the local temperature felt by the Kodama--Unruh--DeWitt detector can be written as 
\beq\label{UH}
T=\sqrt{T^2_U + \alpha T^2_{GH}}\,,
\label{z2}
\eeq 
where $T_U$ is the Unruh temperature associated with the acceleration of the Kodama observer, $T_{GH}$ is the 
Gibbons--Hawking temperature and $\alpha=1$ for the de Sitter space-time, $\alpha =0$ for Minkowski space-time (this is the original Unruh
effect) and $\alpha=-1$ for the ``massless" AdS topological black hole. This formula may help to understand better the relation between the Unruh-like effects and the genuine presence of a thermal bath and shows that, in general, the Kodama--Unruh detector gives an intricate relation between Killing--Hayward temperature and other invariant temperatures such as the Unruh's one. Note that $T$ in Eq.~\eqref{z} is greater than $T_{U}=A(r)/2\pi$ for $r>r_{H}$, where $A$ is the local acceleration of an observer following a Killing trajectory in the black hole space-time, a fact that has been interpreted as a violation of the equivalence principle \cite{Singleton:2011vh}. We prefer to interpret this effect as due to the additional presence of the Hawking radiation over the vacuum thermal Unruh's noise.

\subsection{Finite-time effects in stationary space-times}

We now present a brief discussion of finite-time effects which will be relevant to the following discussion on asymptotic behaviour: how is the thermal distribution of the response function reached in the limit of very large times? 
In the case of non inertial particle detector in Minkowski space-time, see \cite{svaiter}, and for de Sitter FRW  space see \cite{proco}.

To answer this, we consider the finite time contribution due to the fluctuating tail (the $J_\tau$ term in Eq.~\eqref{Lr}) for 
de Sitter or black hole cases compared to the thermal value given by the time-independent part.  A direct calculation of the tail (\ref{L00}) using Eq.~\eqref{figa} for black holes (in particular  Eq.~\eqref{j} for dS) and the fact that
\beq
\mbox{csch}^{2}(x)=4\sum_{n=1}^{\infty}n\,e^{-2nx}
\eeq
gives
\beq\label{tailds}
J_{\tau} &=& \frac{\kappa^2_{l}}{8\pi^2}\,\int_{\Delta \tau}^{\infty}d s\, \frac{\cos (E\,s)}{\sinh^2 \left(\frac{\kappa_{l} s}{2}\right)}\nonumber \\
&=& \frac{E}{2\pi^{2}}\sum_{n=1}^{\infty}\frac{n e^{-2\pi nT_{H}\Delta \tau}}{n^2+E^2/4\pi^{2}T_{H}^{2}} \times \nonumber  \\
& & \hspace{10pt} \times \left(\frac{2\pi T_{H}}{E}
n\cos(E \Delta \tau)-\sin (E \Delta\tau)\right)
\eeq
where $\kappa_l$ is the rescaled surface gravity and $T_H=\kappa_l/2\pi\equiv\kappa/2\pi \sqrt{V}$ the local Hawking temperature.  We recall that $\kappa=H_{0}$ for de Sitter space and $\kappa=V'(r_{H})/2$ for the black hole: these quantities in fact determine the characteristic time-scales the thermalization time has to be compared to.\

We consider as before the peculiar Kodama observer for which $V=1$, so that $T_H=\kappa/2\pi$.  As a general feature, the fluctuating tail term drops out exponentially for large $\Delta\tau$, that is for long proper time intervals in which the detector stays on.  In order to analyze the approaching to an equilibrium condition of the response function, we consider the ratio between the finite-time expression -- the sum of  $\dot{F}$ and the tail $J_\tau$ -- and $\dot{F}$  alone, with the agreement that equilibrium is attained whenever
\begin{equation*}
R_{eq}=\frac{\dot{F} + J_\tau}{\dot{F}} \sim O(1)\;.
\end{equation*}
Looking at \eqref{tailds} one easily sees that the equilibrium value, $R_{eq}=1$, is reached sooner if $E/\kappa \ll 1$.  To be more precise, irrespective of the absolute value of $\kappa$, \textit{a detector that is switched on for a time much shorter than the characteristic time-scale $\Delta\tau\ll\kappa^{-1}$, detects a thermal bath only with particles whose energies are $E \ll \kappa$; on the other hand, a thermal equilibrium for particles with energies $E\gg \kappa$ is registered only if the detector lifetime is $\Delta\tau\gg\kappa^{-1}$}, which is the age of the universe. The Hubble scale corresponds to an extremely small energy scale of order $10^{-42}{\rm Gev}$, therefore $E\gg\kappa$ is the physical region.\

It easy to see that if the factor $V<1$, the thermalization time decreases for every energy scale.

\section{The FRW and asymptotically de Sitter space-times}

To apply the Unruh--DeWitt detector formalism to cosmology we consider a generic FRW spatially flat space-time.
This case has been investigated also in \cite{obadia08} (see also \cite{proco1}).
 Recall that here the areal radius is $R=ra(t)$. Thus, for the Kodama observer, one has
\beq
r(t)=\frac{K}{a(t)},
\eeq
with constant $K$. For a radial trajectory, the proper time in FRW is
\beq
d\tau^2=a(\eta)(d\eta^2-dr^2)\,.
\eeq
As a function of the proper time, the conformal time along a Kodama trajectory is
\beq
\eta(\tau) &=& -\int d \tau \,\frac{1}{a(\eta) \sqrt{1-K^2H^2(\tau)}} \nonumber \\
&\equiv& -\int d\tau\,\frac{1}{a(\tau)\sqrt{V(\tau)}} \,,
\eeq
$H(\tau)$ being the Hubble parameter as a function of proper time. In general, we may use Eq.~\eqref{L6} in which, for radial Kodama observer, one has
\beq
x(\tau)&=&(\eta(\tau), r(\tau),0,0)\nonumber \\
&=&\left(-\int \frac{1}{a(\tau)\sqrt{V(\tau)}} d \tau, \frac{K}{a(\tau)},0,0\right).
\eeq 
As a warm up, we first revisit the  well  known example of FRW space is the stationary flat de Sitter expanding (contracting) space-time, which in the FRW context is defined by considering $a(t)=e^{H_0t}$. Thus, 
\beq
ds^2=-dt^2+e^{2H_0t} (dr^2+r^2d\Omega^2)\,.
\eeq
Here $H(t)=H_0$ is constant as well as $V=V_0=1-H_0^2K^2$. For Kodama observers 
\beq
\tau=\sqrt{V_0}\, t\,, \quad \quad a(\tau)=e^{\frac{H_0}{\sqrt{V_0}}\tau}\,,
\eeq
and
\beq
\eta(\tau)= -\frac{1}{H_0}e^{-\frac{H_0}{\sqrt{V_0}}\tau}\,, \quad r(\tau)= K \,e^{-\frac{H_0}{\sqrt{V_0}}\tau}\,, 
\eeq
thus, the geodesic distance is
\beq\label{tetta}
\sigma_{dS}^2(\tau,s)=-\frac{4\,V_0}{H_0^2}\sinh^2\left(\frac{H_0 \,s}{2\sqrt{V_0}}\right)\,.
\eeq
This result is formally equal to the one obtained for the generic static black hole (\ref{sinh2}). 
Since again $\sigma^2(\tau, s)= \sigma^2(s)=\sigma^2(-s)$, we may use (\ref{Lr}) and obtain, for $E >0$ and in the infinite time limit
\beq\label{desitt}
\frac{d F}{ d\tau}=\frac{H_0}{8\pi^2\sqrt{V_0}}\,\int_{-\infty}^{\infty}d x\, e^{-\frac{2 i \sqrt{V_0} E x}{H_0}}\left[\frac{1}{x^2}-\frac{1}{\sinh^2 x}\right]
\eeq 
Again, we arrive at
\beq 
\frac{d F_{dS}}{ d\tau}=\frac{1}{2\pi} \,\frac{E}{\exp\left(\frac{2\pi \sqrt{V_0} E}{H_0}\right)-1}\,,
\label{jds}
\eeq
which shows again that the Unruh--DeWitt thermometer in the FRW de Sitter space detects a quantum system in thermal equilibrium at a temperature $T=H_0/2\pi \sqrt{V_0}$. Here, the Tolman factor takes the form a Lorentz $\gamma$-factor, which represents the Unruh acceleration part. In fact, we recall that the four-acceleration of a Kodama observer in a FRW space-time has the expression
\beq
A^2 = A^\mu A_\mu = 
K^{2}\left[\frac{\dot H(t) + (1- H^2(t) K^2) H^2(t)}{(1 - H^2(t) K^2)^\frac{3}{2}}\right]^2\, .
\eeq
As a result, for dS space in a time dependent spatially flat patch we have 
\beq
A^2=\frac{K^2H_0^4}{1-K^2H_0^2}\,,
\eeq
showing that 
\[
\frac{H_{0}}{\sqrt{1-H_{0}^{2}K^{2}}}=\sqrt{H_{0}^{2}+A^{2}}\,,
\]
in agreement with the dS static calculation. When $K=0$, that is when the detector is co-moving, one has $V_0=1$ and the classical Gibbons--Hawking result $T_{dS}=H_0/2\pi$ is recovered. \\

Let us come to consider the more realistic scenario of a truly dynamical space-time of cosmological interest.
From previous considerations, our basic formulas for the transition rate of the detector (\ref{L6}) are manageable -- in the sense that we can extract quantitative information -- only in the few highly symmetrical circumstances mentioned in Section 4. As it will be clear at the end of this section, any departure from those models is responsible for significant difficulties. For instance, let us take on the case of homogeneous, spatially flat, universe dominated by cold matter and cosmological constant. The scale factor is (e.g. \cite{muk})
\beq
a(t)=a_{0}\sinh^{2/3}\left(\frac{3}{2}\sqrt{\Omega_{\Lambda}}H_{0}t\right) \label{scale factor}
\eeq
where $a_0=(\Omega_{m}/\Omega_{\Lambda})^{1/3}$ and $\Omega_{m}+\Omega_{\Lambda}=1$; $H_{0}= \sqrt{8 \pi \rho_{cr}/3}$ and $\Omega_A$ represents the relative density of matter (if $A=m$) or cosmological constant(if $A=\Lambda$). Setting $h\equiv \sqrt{\Omega_{\Lambda}}H_{0}$ for simplicity, its current value is of order $h \approx 2 \times 10^{-18} sec^{-1}$. Upon integration the conformal time becomes
\beq
\eta(t)&=&\frac{1}{a_0 h}\left\lbrace\frac{\Gamma\left(\frac{1}{6}\right)\Gamma\left(\frac{4}{3}\right)}{\sqrt{\pi}} -\mbox{sech}^{2/3}\left(\frac{3}{2}h t\right)\times \nonumber \right.\\ & & \left.\hspace{20pt}\times  _{2}F_{1}\left(\frac{5}{6},\frac{1}{3},\frac{4}{3};\mbox{sech}^{2}\left(\frac{3}{2} h t\right)\right)\right\rbrace
\eeq
where $_{2}F_{1}(a,b,c;z)$ is a hypergeometric function and the constant has been opportunely chosen so that at the Big Bang $\eta(t=0)=0$.
The detector's proper time is related to the cosmic time through a manageable expression only if we limit ourselves to consider co-moving detectors: $\tau(t)-\tau_0 = \int dt \sqrt{1-H^2(t) K^2}$ so that for $K=0$, $\Delta \tau(t) = t$, $\Delta \tau$ being the proper time interval during which the detector is turned on.  Unlike the stationary cases analyzed previously, this model presents a Big Bang singularity at the origin of the time coordinate, so that the detector must be switched on at some $\tau_0>0$. In particular, the Big Bang prevents taking the limit as $\tau_{0}\to-\infty$. By the same reason, the scale factor (\ref{scale factor}) is defined only for positive values of the argument: a new feature with respect to what we have seen in the previous Sections. As a consequence, $a(t-s)$ is defined as in (\ref{scale factor}) only for $t-s > 0$ and trivially continued outside the interval in order to make well defined the transition rate \eqref{L6}.\

Leaving the technical details to the Appendix A, we obtain the following response function
\beq\label{integr}
&&\dot F_\tau =\dot F_{dS}+J_{dS,\tau}+\nonumber \\ 
&-&\frac{h^2}{2 \pi^2} \sum_{n=1}^{\infty} \sum_{k=1}^{3n-1}g(n,k)\, e^{-3n\, h\Delta\tau}  \times\nonumber   \\
&&\frac{e^{hk\Delta \tau} \Big(hk \cos(E \Delta\tau) +E \sin(E \Delta\tau)\Big)-h k}{h^2k^2 + E^2} 
\eeq
in which $\dot F_{dS}$ is the De Sitter $\tau$ independent contribution given by Eq. (\ref{jds}) but with effective Hubble constant $h=\sqrt{\Omega_{\Lambda}}H_{0}$ and $ J_{dS, \tau}$ is the related tail given by Eq. (\ref{tailds}).  The numerical coefficients $g(n,k)$ can in principle be computed but enter in a tail which decays exponentially fast in the switching time and which also contain oscillating terms.

We may take the limit $\Delta\tau\rightarrow\infty$ and observe  that every  $\tau$ dependent term of this expression 
goes to zero. This is the main result of our paper.
\
To summarize, we may say that the detector clicks close to a de Sitter response and reaches thermalization (possibly, through decaying oscillations) as  $\Delta \tau$ is sufficiently large. In fact, as far as the regime $h \Delta \tau \gg 1$ is concerned, de Sitter space-time is recovered. We may think of this as describing a de Sitter thermal noise continuously perturbed by the expansion (or contraction) of the universe. In particular, insofar as we can speak of temperature, in this large-time regime the detector registers the de Sitter temperature $h/2\pi$, equal to the large-time limit of the horizon temperature parameter given by the surface gravity, which in the present case has the exact but slow long-time evolution
$$
T_{H}=\frac{h}{2\pi} \left[\coth(3 h t/2) - 3/4\, \mbox{sech}(3 h t/2) \mbox{csch}(3 h t/2)\right]
$$
It is worth noting that, while in the stationary phase $h \Delta \tau \gg 1$ the limiting result is consistent with the limiting value of the surface gravity, in the non-stationary regime it seems less trivial to compare the results of the two methods, because it has not been possible to extract a temperature parameter from the transition rate of the detector, but asymptotically.

\section{Conclusions}

In this paper,  with the aim to better understand the temperature-versus-surface gravity paradigm, the asymptotic results obtained by semiclassical method in previous papers  have been tested with more reliable quantum field theory techniques as the Unruh--DeWitt detector analysis. For black holes and pure de Sitter space the two analysis are mutually consistent and even predict the dependence of the temperature on position or acceleration. Moreover, the analysis of the oscillating tail has been extended to stationary black holes.\

For cosmology and away from de Sitter space the thermal interpretation, strictly speaking, is lost but the detector still gets excited by the expansion of the universe. By accepting the surface gravity versus temperature paradigm we would expect a quasi-thermal excitation rate of the form
$$
\dot{F}\sim E\exp(-E/T_{H}(t))
$$
$T_{H}(t)$ being given by our last expression above. That is, although in a generic FRW  space-time the thermal interpretation breaks down in most of the cases because of the time-dependence of the background, still this time dependence of the transition rate could be expected to mainly reside in an effective temperature parameter. But using a comoving detector this is not what we have found.  
For instance, in the Einstein-de Sitter regime there seems to be excitations of non-thermal type and we showed that the scale factor of the flat $\Lambda$CDM-cosmology has no other temperatures in action than the de Sitter one.\\

It remains to see whether there is any non trivial quasi-thermal effect on accelerated, or more general, Kodama trajectories. In the affirmative case, that would mean that the horizon surface gravity and temperature should be associated more likely to vacuum correlations than to particle creation and forces,  in our view, a different interpretation of the tunneling picture.  In this respect, the classical Parker's papers on particle creation \cite{Parker:1968mv,Parker:1969au} are certainly relevant. One possibility is that the horizon surface gravity could represent an intrinsic property of the horizon itself, leading to some kind of holographic description, while the detector in the bulk simply clicks because it is embedded in a changing geometry. In fact, we would expect the clicks in almost any changing geometry, even for those lacking a trapping horizon.

\ack{We thank S. A. Hayward and G. Cognola for useful discussions.}\\

\renewcommand{\theequation}{A-\arabic{equation}}
\setcounter{equation}{0}
\section*{APPENDIX A: Response function for $\Lambda$CDM model}

In order to obtain \eqref{integr}, we define the variable $x=\exp\left( -\frac{3}{2}h \Delta\tau \right)$ and expand the inverse $\sigma^2(x,s)$, given by the scale factor (\ref{scale factor}), around $x=0$ (\textit{i.e.} $\Delta\tau\rightarrow\infty$).  We obtain a reasonably simple expansion in even powers of $x$ given by

\begin{equation}
\frac{1}{\sigma^2(x,s)} =\frac{1}{\sigma_{dS}^2(s)} -h^2\, \sum_{n=1}^{\infty} \left( x^{2n}\, \sum_{k=1}^{3n-1}\, g(n,k)\, e^{k\, h s} \right)\,,
\label{vanzo}
\end{equation}
On the right hand side, the first term is the constant term of the expansion and happens to be the pure de Sitter contribution, i.e.
\beq
\sigma_{dS}^2(s)=-4h^{-2}\sinh^{2}\left(\frac{h s}{2}\right)
\eeq
with the effective Hubble constant $h=\sqrt{\Omega_{\Lambda}}H_{0}$.  Numerical hints given by the coefficients of the expansion up to the 10$^{th}$ order in $x$, allow us to make a conjecture that the $g(n,k)$'s in the second term have a mean decreasing behavior and are bounded in the interval $(0,1)$, but the main point is that the series in eq.(\ref{vanzo}) is absolutely convergent with a finite radius of convergence which includes any $t>0$, namely the entire range of integration. 

Hence, integrating term by term the expression (\ref{vanzo}), and making use of (\ref{L6}), 
for finite $\Delta \tau$ one has eq.(\ref{integr}).

In the $\Delta \tau \rightarrow \infty$ limit, we can focus on the leading exponentials contained in the last term of this expression: these are the $k=(3n-1)$ terms, which are all dominated by a common factor $\exp\left( - h k \Delta \tau \right)$.  All the other terms are even more damped, so the convergence to zero is evident.


\begin{thebibliography}{99}


\bibitem{Haw}S W Hawking, Nature {\bf 248}, 30 (1974); {Commun. Math. Phys.}{\bf 43}, 199 (1975) [Erratum-ibid.\ {\bf 46}, 206 (1976)]
\bibitem{dewitt}
B.~S.~DeWitt,
Phys.\ Rept.\ {\bf 19} (1975) 295.
\bibitem{BD}N D Birrell \& P C W Davies, Quantum fields in curved space
(Cambridge University Press 1982).
{\bf 43}, 199 (1975) [Erratum-ibid.\ {\bf 46}, 206 (1976)]
\bibitem{wald}R.~M.~Wald, Quantum Field Theory in Curved Spacetime and
Black Hole Thermodynamics (Chicago Lectures in Physics, Chicago University Press 1994).
\bibitem{fulling} S. A. Fulling, Aspects of Quantum Field Theory in Curved Space-time (Cambridge University Press 1996).
\bibitem{igor}
V.~P.~Frolov and I.~D.~Novikov, Black hole physics, Kluwer Academic
Publisher, 2007.

\bibitem{Unruh:2008zz}
 W.~G.~Unruh,
 Phil.\ Trans.\ Roy.\ Soc.\ Lond.\  A {\bf 366}, 2905 (2008).

\bibitem{Barcelo:2005fc}
 C.~Barcelo, S.~Liberati, M.~Visser,
 Living Rev.\ Rel.\  {\bf 8}, 12 (2005).
 [gr-qc/0505065].

\bibitem{kodama}
H.~Kodama,
Prog.\ Theor.\ Phys.\ {\bf 63}, 1217 (1980).

\bibitem{bob07}
R.~Di Criscienzo, M.~Nadalini, L.~Vanzo, S.~Zerbini and G.~Zoccatelli,
Phys.\ Lett.\ {\bf B657}, 107 (2007).

\bibitem{sean09}
S.~A.~Hayward, R.~Di Criscienzo, L.~Vanzo, M.~Nadalini and S.~Zerbini,
Class.\ Quant.\ Grav.\ {\bf 26 }, 062001 (2009).

\bibitem{bob09}
R.~Di Criscienzo, S.~A.~Hayward, M.~Nadalini, L.~Vanzo and S.~Zerbini,
Class.\ Quant.\ Grav.\ {\bf 27}, 015006 (2010).

\bibitem{av}
G. Abreu and M. Visser, {\it Phys. Rev.} {\bf D} 82, 044027 (2010)

\bibitem{tr}
L.~Vanzo, G.~Acquaviva, R.~Di Criscienzo,
``Tunnelling Methods and Hawking's radiation: achievements and prospects,''
to appear in CGQ special issue (2011). 
 [arXiv:1106.4153 [gr-qc]].


\bibitem{PW}
M.~K.~Parikh and F.~Wilczek,
Phys.\ Rev.\ Lett.\ {\bf 85}, 5042 (2000).

\bibitem{visser}
M.~Visser,
Int.\ J.\ Mod.\ Phys.\ {\bf D12}, 649 (2003);
A.~B.~Nielsen and M.~Visser,
Class.\ Quant.\ Grav.\ {\bf 23}, 4637 (2006).

\bibitem{Angh}
M. Angheben, M. Nadalini, L. Vanzo and S. Zerbini,
JHEP {\bf 0505}, 014 (2005);
M. Nadalini, L. Vanzo and S. Zerbini,
J. Physics A: Math. Gen. {\bf 39}, 6601 (2006).

\bibitem{tanu}
K.~Srinivasan and
T.~Padmanabhan, 
Phys.~Rev.~{\bf D 60}, 24007 (1999).

\bibitem{mann}
R.~Kerner and R.~B.~Mann,
Phys.\ Rev.\ D {\bf 73}, 104010 (2006)

\bibitem{obr}
A.~J.~M.~Medved and E.~C.~Vagenas,
Mod.\ Phys.\ Lett.\ A {\bf 20}, 2449 (2005);
M.~Arzano, A.~J.~M.~Medved and E.~C.~Vagenas,
JHEP {\bf 0509}, 037 (2005); R.~Banerjee and B.~R.~Majhi,
Phys.\ Lett.\ B {\bf 662}, 62 (2008).

\bibitem{bob08}
R.~Di Criscienzo and L.~Vanzo,
Europhys.\ Lett.\ {\bf 82}, 60001 (2008). 

\bibitem{bob10}
R.~Di Criscienzo, L.~Vanzo and S.~Zerbini,
JHEP {\bf 1005}, 092 (2010).

\bibitem{Lindesay:2010uv}
J.~Lindesay and P.~Sheldon,
Class.\ Quant.\ Grav.\ {\bf 27}, 215015 (2010).

\bibitem{Brown:2009by}
B.~A.~Brown and J.~Lindesay,
AIP Conf.\ Proc.\ {\bf 1280}, 3 (2010)
[arXiv:0904.4192 [gr-qc]].

\bibitem{seanrep}
S.~A.~Hayward, R.~Di Criscienzo, M.~Nadalini, L.~Vanzo and S.~Zerbini,
arXiv:0909.2956 [gr-qc]. 

\bibitem{D}
 E.~T.~Akhmedov, V.~Akhmedova, D.~Singleton,
 Phys.\ Lett.\  {\bf B642}, 124-128 (2006);
 E.~T.~Akhmedov, V.~Akhmedova, T.~Pilling {\it et al.},
 Int.\ J.\ Mod.\ Phys.\  {\bf A22}, 1705-1715 (2007);
 E.~T.~Akhmedov, T.~Pilling, D.~Singleton,
 Int.\ J.\ Mod.\ Phys.\  {\bf D17}, 2453-2458 (2008);
 V.~Akhmedova, T.~Pilling, A.~de Gill {\it et al.},
 Phys.\ Lett.\  {\bf B666}, 269-271 (2008).


\bibitem{sean}
S.~A.~Hayward,
Class.\ Quant.\ Grav.\ {\bf 15}, 3147 (1998).


\bibitem{peng}
J.~Peng and S.~A.~Hayward, 
``Cosmological Hawking Radiation''
Center for Astrophysics, Shanghai Normal University preprint (2010).


\bibitem{Wu:2008ir}
S.~F.~Wu, B.~Wang, G.~H.~Yang and P.~M.~Zhang,
Class.\ Quant.\ Grav.\ {\bf 25}, 235018 (2008).


\bibitem{chen}
Y.~X.~Chen, J.~L.~Li, and Y.~Q.~Wang,
arXiv:1008.3215 [hep-th].


\bibitem{Brout:1987tq}
R.~Brout, G.~Horwitz and D.~Weil,
Phys.\ Lett.\ B {\bf 192}, 318 (1987).

\bibitem{ugo}
J.~Bros, H.~Epstein and U.~Moschella,
JCAP {\bf 0802}, 003 (2008) ;
J.~Bros, H.~Epstein and U.~Moschella,
Annales Henri Poincare {\bf 11}, 611 (2010);
J.~Bros, H.~Epstein, M.~Gaudin, U.~Moschella and V.~Pasquier,
Commun.\ Math.\ Phys.\ {\bf 295}, 261 (2010).

\bibitem{Volo}
G.~E.~Volovik, JETP Lett.\ {\bf 90}, 1 (2009).


\bibitem{crispino}
L.~C.~B.~Crispino, A.~Higuchi and G.~E.~A.~Matsas,
Rev.\ Mod.\ Phys.\ {\bf 80} (2008) 787.

\bibitem{casadio}
R.~Casadio, S.~Chiodini, A.~Orlandi, G.~Acquaviva, R.~Di Criscienzo and L.~Vanzo,
arXiv:1011.3336 [gr-qc].

\bibitem{Kothawala:2009aj}
 D.~Kothawala and T.~Padmanabhan,
 Phys.\ Lett.\  B {\bf 690} (2010) 201
 [arXiv:0911.1017 [gr-qc]].
\bibitem{obadia08}
N.~Obadia,
Phys.\ Rev.\ D {\bf 78}, 083532 (2008).


\bibitem{nico}
V.~Moretti and N.~Pinamonti,
``State independence for tunneling processes through black hole horizons and
Hawking radiation,'', to appear in Comm. Math. Phys,
arXiv:1011.2994 [gr-qc].

\bibitem{T}
S.~Takagi, Prog.\ Theoretical Phys. Supp\ Grav.\ {\bf 88} (2004) 1.

\bibitem{S}
S.~Schlicht,
Class.\ Quant.\ Grav.\ {\bf 21} (2004) 4647.

%


\bibitem{langlois}
P.~Langlois,
Annals Phys.\ {\bf 321} (2006) 2027.

\bibitem{OM}
N.~Obadia and M.~Milgrom,
Phys.\ Rev.\ D {\bf 75} (2007) 065006.

\bibitem{louko}
J.~Louko and A.~Satz,
Class.\ Quant.\ Grav.\ {\bf 23} (2006) 6321; 
J.~Louko and A.~Satz,
Class.\ Quant.\ Grav.\ {\bf 25}, 055012 (2008).

\bibitem{GH}
G.~W.~Gibbons and S.~W~Hawking,
Phys.\ Rev.\ D {\bf 14}, 2738 (1977).

\bibitem{thirr}
H.~Narnhofer, I.~Peter and W.~E.~Thirring,
Int.\ J.\ Mod.\ Phys.\ B {\bf 10}, 1507 Int.~(1996).

\bibitem{deser98}
S.~Deser and O.~Levin,
Class.\ Quant.\ Grav.\ {\bf 14} (1997) L163.

\bibitem{Singleton:2011vh}
  D.~Singleton, S.~Wilburn,
  Phys.\ Rev.\ Lett.\  {\bf 107 } (2011)  081102.
  [arXiv:1102.5564 [gr-qc]].

\bibitem{Beng} S. {\AA}minneborg, I. Bengtsson,
S. Holst and P. Peld\'{a}n,
Class. Quantum Grav. 13 (1996) 2707.

\bibitem{Mann} R.B. Mann,
Class. Quantum Grav. 14 (1997) L109.

\bibitem{Brill} D.R. Brill,
Helv. Phys. Acta 69 (1996) 249;
D.R. Brill, J. Louko and P. Peld\'{a}n,
Phys. Rev. D.56 (1997) 3600.

\bibitem{Vanzo} L. Vanzo, 
Phys. Rev. D56 (1997) 6475.

\bibitem{svaiter}
  B.~F.~Svaiter, N.~F.~Svaiter,
  Phys.\ Rev.\  {\bf D 46}, 5267-5277 (1992).
  
\bibitem{proco}
B. Garbrecht and T. Prokopec, Class.~Quant.~Grav.~{\bf 21}, 4993 (2004).
  
\bibitem{proco1}
B. Garbrecht and T. Prokopec, Phys.~Rev.~{\bf D 70}, 083529 (2004).


\bibitem{muk}
V. Mukhanov, {\it Physical Foundations of Cosmology} (Cambridge: Cambridge University Press 2005)

\bibitem{Parker:1968mv}
  L.~Parker,
  Phys.\ Rev.\ Lett.\  {\bf 21 } (1968)  562-564.
 
\bibitem{Parker:1969au}
  L.~Parker,
  Phys.\ Rev.\  {\bf 183 } (1969)  1057-1068.
 
\end{thebibliography}
\end{document}